\documentclass[preprint]{aastex}

\def\msun{{\rm\,M_\odot}}
\newcommand{\etal}{et al.\ }
\newcommand{\kms}{\, {\rm km\, s}^{-1}}

\newcommand{\mpc}{\, {\rm Mpc}}

\newcommand{\hmpc}{\, h^{-1} \mpc}

\newcommand{\lya}{Ly$\alpha$ }
\newcommand{\lyaf}{Ly$\alpha$ forest}

\newcommand{\bF}{\bar{F}}
\newcommand{\hi}{\mbox{H\,{\scriptsize I}\ }}

\slugcomment{Submitted to ApJL}

\begin{document}

\title{Evolution of the Ionizing Radiation Background and 
Star Formation in the Aftermath of Cosmological Reionization}

\author{Renyue Cen\altaffilmark{1} and Patrick McDonald\altaffilmark{2}}

\altaffiltext{1} {Princeton University Observatory, 
Princeton University, Princeton, NJ 08544; cen@astro.princeton.edu}

\altaffiltext{2} {Department of Physics, 
Princeton University, Princeton, NJ 08544; pmcdonal@feynman.princeton.edu}

\accepted{ }

\begin{abstract}

The temporal evolution of the ionizing UV background radiation 
field at high redshift provides a probe of the evolution of the early 
star formation rate.
By comparing the observed levels of absorption in the highest 
redshift quasar spectra to the predictions of a hydrodynamic 
simulation, we determine the evolution of the photoionization 
rate ($\Gamma$) for neutral hydrogen in the intergalactic medium, 
over the redshift range $4.9\lesssim z \lesssim 6.1$.  
After accounting for sampling variance, we infer a sharp increase in 
$\Gamma$ from $z\simeq 6.1$ to $z\simeq 5.8$, 
probably implying reionization at this redshift.  We find 
a {\it decrease} in $\Gamma$ from $z\simeq 5.6$ 
to $5.2$, at $3\sigma$ significance.
This feature is a generic signature in the aftermath of 
reionization, entirely consistent with 
the cosmological reionization process being completed at $z\sim 6.1$.

This generic feature is a result of a significant change
in the star formation rate subsequent to the cosmological reionization.
There is an abrupt rise of the temperature of the intergalactic medium
due to photo-heating, when it is reionized.
This translates to a correspondingly sudden jump in the Jeans mass and
a dramatic suppression of gas accretion onto the most
abundant (sub-galactic) halos at the epochs of interest.
The star formation rate suffers a temporary setback in the aftermath
of reionization, resulting in a temporary decrease in 
the amplitude of the ionizing radiation field.

\end{abstract}


\section{Introduction}

The conventional wisdom based on the standard theory of structure
formation says that
the universe was reionized sometime in the redshift range 
$z=6-12$ (Barkana \& Loeb 2001).
The relatively large uncertain range in redshift
reflects our imperfect knowledge
of the density fluctuations on small scales,
star formation processes (e.g., efficiency, IMF, etc.)
and feedback processes at high redshift.
The latest observations of high redshift quasars 
(e.g., Fan \etal 2001) are beginning to probe the lower
bound of that window
and suggestions have been made that
we may be witnessing the end phase of the cosmological
reionization process at $z\sim 6$, based solely on 
the appearance of a precipitous drop of transmitted flux
at the rest-frame \lya wavelength 
near $z\sim 6$ in a single quasar
spectrum (Becker \etal 2001, hereafter B01; Barkana 2001).

In this {\it Letter} we present an analysis of the 
ionizing background radiation field in the redshift range $z=4.9-6.1$,
using the absorption measurements in B01.
Combining with previous data at lower redshift,
we find that, coming from high redshift,
the ionizing radiation intensity 
displays a sharp rise at $z\sim 6$ peaking at $z=5.6$,
a significant downturn from $z\sim 5.6$ to $5.2$ by a factor of $\sim 0.6$,
and subsequently a consistent 
ascent from $z=5.0$ to $z=2.4$ by a factor of $\sim 4$. 
All these three features 
are consistent with a picture that
the cosmological reionization was near completion at $z\sim 6$.

The first feature (i.e., the initial sharp rise)
has been predicted by 
several authors previously 
(Cen \& Ostriker 1993;
Gnedin 2000a; Miralda-Escud\'e, Haehnelt, \& Rees 2001).
At present,
the primary uncertainty for its determination observationally
is the possibility that the high level of absorption in the single
observed quasar at $z=6.28$ is some kind of anomaly (we
show that it cannot be a simple statistical fluctuation in
the absorption level).
Additional observed quasars at $z\gtrsim 6.3$ will be critical in this
regard.
Here we focus our attention on the relatively more robust
measurements of the ionizing radiation field at $z<6$ 
and show that the observed
pause in the rise of the 
amplitude of the ionizing radiation field 
from $z\sim 5.6$ to $5.0$ 
could be due to suppression of star formation following
reionization at $z\sim 6$.

\section{Reionization and the Evolution of the Ionizing Background}

Assuming the model in which the IGM is almost completely ionized
and the radiation background is uniform, we infer the 
evolution of the \hi ionization rate by requiring that the mean 
flux decrement in simulated spectra matches the observed values
in the upper panel of Figure 2 of B01.  
This method was pioneered by Rauch 
et al. (1997) and extended in McDonald et al. (2000,2001)
and McDonald \& Miralda-Escud\'e (2001).  
We apply the standard procedure using a hydrodynamic simulation
of a flat universe dominated by a cosmological constant and
cold dark matter
($\Lambda$CDM), with CDM density $\Omega_m=0.3$, Hubble parameter
$h=0.67$ ($H_0=100~h {\rm \kms/Mpc}$),
baryon density $\Omega_b=0.035$, power spectrum normalization
$\sigma_8=0.9$, and large-scale primordial power spectrum slope $n=1$.
The simulation is Eulerian, with box size $25\hmpc$ 
divided into $768^3$ cells
(see Cen \etal 2001 for a more complete description).  
We use outputs from the simulation at $z=5$ and $z=6$, interpolating
linearly between them when necessary.
We start by assuming the
model in the simulation correctly represents the universe, and then
consider whether or not our results are self-consistent.

\begin{figure}
\plotone{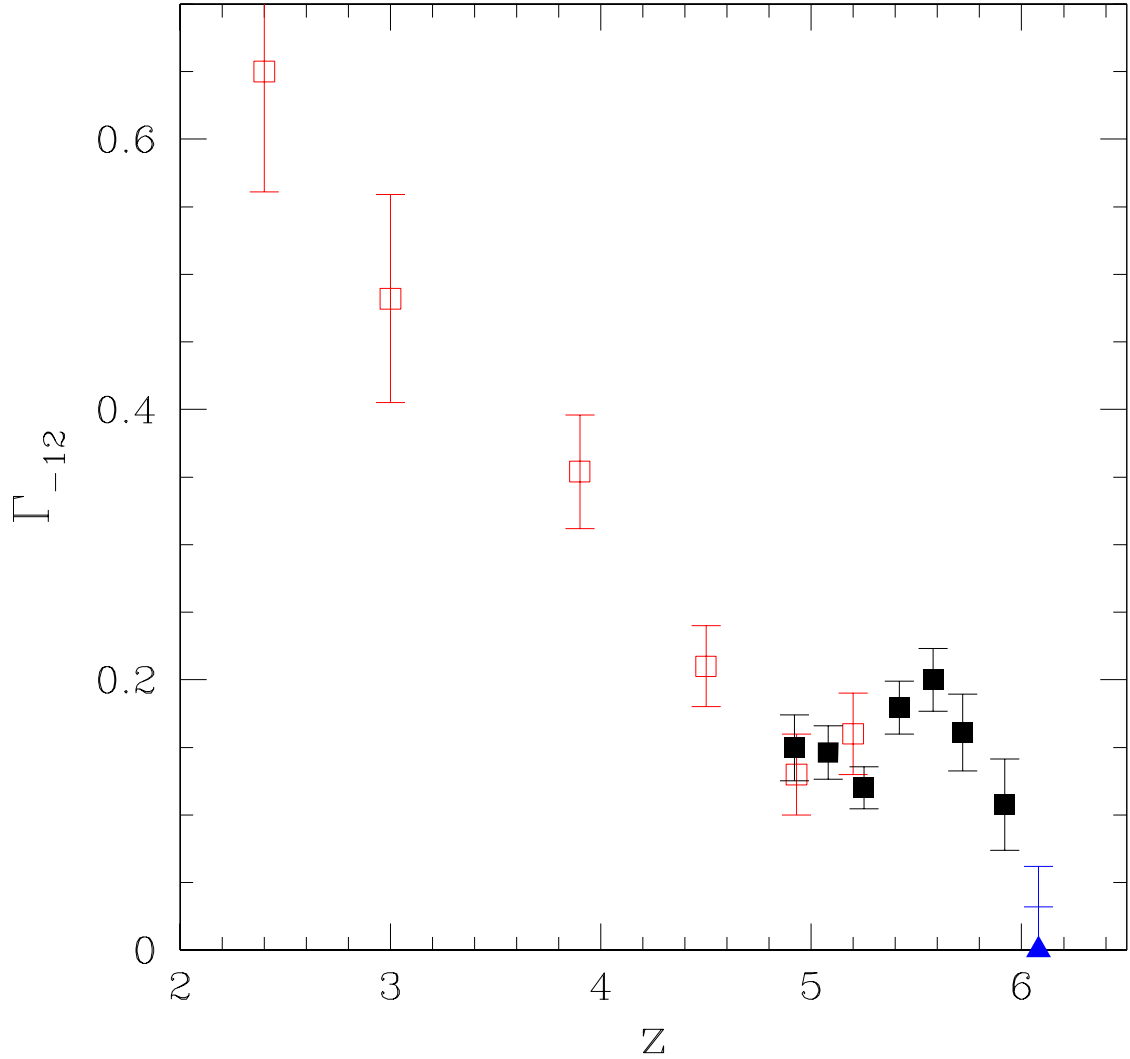}
\caption{Inferred ionization rate as a function of redshift. 
The open squares are from McDonald \& Miralda-Escud\'e (2001)
while the filled squares and triangle are the new results in this paper.
The triangle at the highest redshift is a non-detection with the
error bars indicating 1 and 2$\sigma$ upper limits.  Changing the
cosmological parameters can cause coherent shifts in the 
overall amplitude of the points (not accounted for in the error 
bars), as can changing the temperature
of the gas; however, none of these should cause variation in a
small redshift interval, except if the temperature changed 
suddenly. 
}
\label{Gev}
\end{figure}

Figure \ref{Gev} shows the results for the evolution of $\Gamma_{-12}(z)$,
the hydrogen ionization rate in units of $10^{-12}~{\rm s}^{-1}$.
The open squares are the old results from McDonald \& Miralda-Escud\'e
(2001).  The filled squares are the new results, leaving out the highest
redshift point,  
which we analyze differently than the others because 
the detected transmission is not significant and the detector noise
needs to be taken into account carefully
(this last point is represented by the 
triangle with 1 and 2 $\sigma$ upper limit bars).
The equally spaced bins in redshift are
chosen to coincide with the majority of the points in
Figure 2 of B01.  As discussed below, the results have
been rescaled to apply for the value of the baryon density
favored by big bang nucleosynthesis,
$\Omega_b h^2=0.02$ 
(Burles \& Tytler 1998), instead of the value in the simulation,
$\Omega_b h^2=0.0157$.

In detail, we determine a value of $\Gamma_{-12}$ and its error for
each B01 point using the simulation as follows:
First, we construct the optical depth field, $\tau(v)$ (i.e., 
the spectrum is $F(v)=\exp[-\tau(v)]$)
along a large number of lines of sight through the simulation 
cube ($256^2$ for each face),
normalized using an arbitrary value of $\Gamma_{-12}$
(i.e., $\tau \propto \Gamma^{-1}$ for low density gas in ionization
equilibrium).
We then randomly choose many sets of five $12.5 \hmpc$-long segments of 
spectrum, so that each set forms a combined spectrum of the 
length of our redshift bins,
$\Delta z=0.167$ (note that this length depends slightly on cosmology
and redshift, i.e., the spectra should be shorter by 26\% at $z=6$ in
our model, corresponding to a 13\% increase in the estimated error 
bars, which we account for in the plot).  For each $62.5 \hmpc$ chunk,
we determine the value of $\Gamma_{-12}$ that produces the observed 
value of the mean transmitted flux fraction, $\bF(z)$ [$T(z)$ in 
the notation of B01].  The mean of
the $\Gamma_{-12}$ values for all the chunks is the inferred value of
$\Gamma_{-12}$ for this observed data point, and the dispersion is the 
estimated error.  Most redshift bins in B01 have several measured 
values of $\bF$ in them, so we combine the inferred 
values of $\Gamma_{-12}$ within the bin
using an error weighted average which takes into account fractional 
overlap when necessary (because of the overlapping points, our error
bars are correlated, in the sense that point-to-point differences are
more significant than the error bars seem to indicate).

To test that the treatment of $12.5 \hmpc$ segments as independent 
does not underestimate the error by ignoring correlation along the 
lines of sight on larger
scales, we repeat this procedure using $6.25 \hmpc$ segments, and find
almost identical error bars.

We now turn to the highest redshift point in B01.
The given value and error are
$\bF=0.0038 \pm 0.0026$; however, B01 state that the sky subtraction
has an uncertainty of order 1$\sigma$, so we use $\sigma = 0.0052$ 
when interpreting this point.  B01 also give the measured 
transmitted flux fraction in the region of the spectrum corresponding
to Ly$\beta$ absorption by gas at this redshift, 
$\bF=-0.00024\pm 0.0024$, and point out that
the implied constraint on the \lya optical depth is actually
stronger than the direct one.  We assume the same sky subtraction 
error for this point as for the \lya region.     

We start by analyzing the \lya region.  We use the method described
above to create chunks of spectra of the appropriate length. 
For each value of $\Gamma_{-12}$ we compute $\bF$ for many different
chunks, and compute the mean likelihood of producing 
$\bF=0.0038$ given the error $\sigma=0.0052$.  The relative likelihood
function peaks at $\Gamma_{-12}=0.059$, but does not decrease significantly 
for arbitrarily small $\Gamma_{-12}$.  The likelihood has dropped
by factors of 0.61 and 0.14 at $\Gamma_{-12}=0.096$ and 0.135, 
respectively, so we call these the 1 and 2$\sigma$ upper limits.  

We analyze the Ly$\beta$ region in two ways.  First we repeat the 
above procedure assuming
a constant factor 0.13 suppression of the continuum by \lya\ absorption
corresponding to gas at $z \simeq 5$ (see the lowest redshift point 
in B01).  The inferred 1 and 2$\sigma$ upper limits on $\Gamma_{-12}$
are 0.032 and 0.056.  These constraints are stronger than the ones
from the \lya\ region; however, they are not as realistic as 
possible because they ignore fluctuations in the \lya\ absorption 
in the Ly$\beta$ region, which might either weaken the constraints,
by completely obscuring substantial portions of the Ly$\beta$ 
region, or strengthen the constraints, by providing regions of 
significantly less than average absorption where residual transmission
through the Ly$\beta$ forest could be observed.  We simulate this 
effect by creating \lya\ absorption spectra to serve as the continua
for the Ly$\beta$ region, using the $z=5$ simulation output and
our inferred value of $\Gamma_{-12}$ from the lowest redshift B01 point.
This procedure results in essentially identical constraints to the
assumption of a flat continuum:  $\Gamma_{-12}<0.032$ and 0.062 at
1 and 2$\sigma$, respectively.

It may be possible to obtain even stronger constraints by using 
Ly$\gamma$, as suggested by B01 (X. Fan, private communication), 
or by considering the limits 
imposed by individual pixel values (D. Weinberg, private communication).
However, if reionization is on-going or just completed at $z=6.1$,
the detailed constraints on $\Gamma_{-12}$ obtained 
from any kind of analysis based on a simulation in which reionization 
happened much earlier probably are not very meaningful  
($\Gamma_{-12}$ is not well defined if the radiation background
is inhomogeneous).  Our formal constraints on 
$\Gamma_{-12}$ near $z=6$ should be taken simply as an indication
that something dramatic is happening at this redshift 
and we think it likely indicates the end of the cosmological 
reionization process.

The overall amplitude of our $\Gamma$ results (but not the 
evolution over a short redshift interval) is sensitive
to the cosmological model and gas temperature-density relation
in the simulation,
as one can see from the equation for optical depth at a point
in real space
(i.e., ignoring peculiar velocities and thermal broadening)
with baryon overdensity $\Delta = \rho_b/\bar{\rho_b}$ and
gas temperature $T$:
\begin{equation}
\tau \propto
\frac{\left(\Omega_b h^2\right)^2~\Delta^2}
{T^{0.7}~ H(z)~ \Gamma}~,
\label{taueq}
\end{equation}
where $H(z)$ is the Hubble parameter at 
redshift $z$
[$H(z) \simeq 100~h~ \Omega_m^{1/2}(1+z)^{3/2}$ at high $z$ in a
$\Lambda$CDM universe].  Sensitivity of the inferred value of 
$\Gamma$ to the power spectrum of 
density perturbations enters through the $\Delta^2$ term, as
well as the peculiar velocities, and additional sensitivity to
temperature can arise from the thermal broadening.  The effect of
changing $\Omega_b h^2$ or $H(z)$ on 
the inferred value of $\Gamma$ is easy to correct for, but the
effect of changing the temperature-density relation or the input
power spectrum generally must be investigated by running additional
simulations.  We do this using approximate Hydro-PM simulations 
(hereafter HPM, Gnedin \& Hui 1998), after checking that they give 
similar results
to our fully hydrodynamic simulation (to $\sim 10$\% in $\Gamma$) 
when performed using the same model and initial conditions.  
First, we analyze HPM simulations with a variety of different 
temperature-density relations and find that the thermal broadening
is not very important to the derivation of $\Gamma$, so the dependence
$\Gamma \propto T^{-0.7}$ is accurate (to $\sim \pm 0.1$ in the power
law index), as long as the temperature is specified at the most 
relevant density for the transmission features that determine
$\Gamma$, which we find to be $\Delta \sim 0.35$.  
The temperature in our hydrodynamic
simulation at this density is $5400$ K, but in the aftermath of
reionization the temperature of the gas
would be higher than this
(e.g., Miralda-Escud\'e \& Rees 1994), requiring a correction.  
The amplitude of the 
initial density fluctuations in our hydrodynamic 
simulation is higher by a factor $\sim 1.5$ on the
scale of the \lyaf\ than the measurement of McDonald et al. (2000),
which was independently confirmed by Croft et al. (2001).  
By varying the amplitude, $A$,
in HPM simulations, we determine that 
$\Gamma \propto A^{-1.5}$ (roughly).  The required ionizing
background decreases when $A$ increases because the voids 
where transmission is observed become deeper (i.e., $\Delta$ 
becomes smaller).  In the end, 
the corrections for power spectrum amplitude and temperature
may roughly cancel (if we use $T\sim 10000$ K), so the 
points in Figure \ref{Gev} should
be reasonably accurate; however, we emphasize that the temperature
at this redshift is unknown, so significant uncertainty
in $\Gamma$ remains.

The reader may be wondering at this point ``could the inferred
decrease in $\Gamma_{-12}$ from $z\simeq 5.6$ to $z\simeq 5.2$
possibly be real''? It appears to be a 3$\sigma$ effect and
we think it could be real.  
The same trend is 
clearly visible to the eye in the bottom panel of Figure 2 of
B01, once we consider that the value of their $\tau_{eff}$ is
generally expected to increase with time.  The signal is almost
equally strong in the spectra of two different quasars.
These two are at almost identical redshift so a continuum-related 
effect might be the same in both; however, the observed $\sim 30$\% 
increase in transmission fraction between $z\simeq 5.2$ and
$z\simeq 5.6$ would
require an increase $\lambda^{\sim 4}$ in the quasar continuum.
Finally, we have every reason to believe that our statistical error 
calculation is correct.  The scatter in the inferred values of
$\Gamma_{-12}$ for different B01 points in each redshift bin is
in perfect agreement with our estimated error bars 
(e.g., $\chi^2=13.0$ for 10 degrees of freedom when we 
treat the three average values of $\Gamma_{-12}$ for the bins
in the range $5.2\lesssim z \lesssim 5.6$ as fitting parameters
to be estimated from the individual points with error bars).
Note that, while it may look small to the eye, the fractional error 
on the $z=5.25$ point is actually
larger than the fractional errors on the following two points.
Purely adiabatic temperature evolution $\propto (1+z)^2$,
which constitutes the steepest possible evolution in temperature
(by ignoring the subsequent photo-heating),
instead of the essentially constant temperature in the simulation, would
only cause a 7.5\% relative change in the values of the 
$\Gamma$s.

 We think that the sharp rise of the radiation intensity
 at $z\sim 6.1$ may be a fairly strong case for the completion
 of cosmological reionization at that epoch.
 %However, one is hampered by the finite noises in the small detected
 %fluxes.
 However, note that a value of $\Gamma_{-12}\sim 0.03$ is not 
 inconsistent in
 principle with a fully ionized IGM, since the neutral fraction near
 the mean density is still only $\sim 0.0007$.
 One is hampered by the high optical depth of an IGM that is still
 mostly ionized.
 It will be extremely valuable to look for additional
 signatures of cosmological reionization in its aftermath,
 where more flux is transmitted.
 %and signal-to-noise
 %is substantially improved.
 We will therefore focus our discussion on
 the radiation field at $z\le 6$.

 \section{Discussion}

 In the standard hierarchical structure formation theory
 smaller structures began to form earlier than larger structures,
 simply because the amplitude of density fluctuations 
 is a decreasing function of scale and in linear theory 
 density fluctuations on all scales grow at the same rate.
 Under this general picture
 the universe is conventionally 
 thought to be reionized mostly by photons
 from stellar systems more massive 
 than $10^8\msun$, where
 atomic line cooling provides an efficient energy sink
 for gas that is collected and heated during gravitational collapse
 of halos (Haiman, Rees \& Loeb 1997; Gnedin \& Ostriker 1997).
 Less massive systems, after having produced an insufficient amount 
 of stars (Pop III) to reionize the universe at an earlier epoch,
 can no longer form stars due to lack 
 of cooling processes (Haiman, Thoul, \& Loeb 1996;
 Haiman, Rees \& Loeb 1997; Tegmark \etal 1997).

 After reionization, the temperature of the intergalactic gas
 is raised to $\ge 10^4$K (Cen \& Ostriker 1993;
 Miralda-Escud\'e \& Rees 1994; Gnedin \& Ostriker 1997).
 The subsequent effect of suppression of gas accretion onto 
 small halos has been noted by many authors 
 (Efstathiou 1992;
 Thoul \& Weinberg 1996;  %$\le 30~$km/s
 Quinn, Katz, \& Efstathiou 1996; %$\le 37~$km/s
 Kepner, Babul, \& Spergel 1997;
 Navarro \& Steinmetz 1997; %reduced up to 50 percent in 80-200km/s systems.
 Kitayama \& Ikeuchi 2000;
 Gnedin 2000b). 
 The extent of this effect somewhat varies among the studies
 and it is clear that our current treatment 
 of the reionization process is far from ideal.
 Obviously, a sufficiently adequate study would require simulations
 that include
 an accurate treatment of three-dimensional radiative transfer process
 and have high enough spatial and mass resolution.
 We do not have simulations of this caliber at present
 and attempt to only make a semi-quantitative calculation to 
 illustrate a consequence of reionization on star formation
 and the background radiation field, which we believe
 captures a primary feature of this process
 in a simple and intuitive fashion.

 \begin{figure}
 \plotone{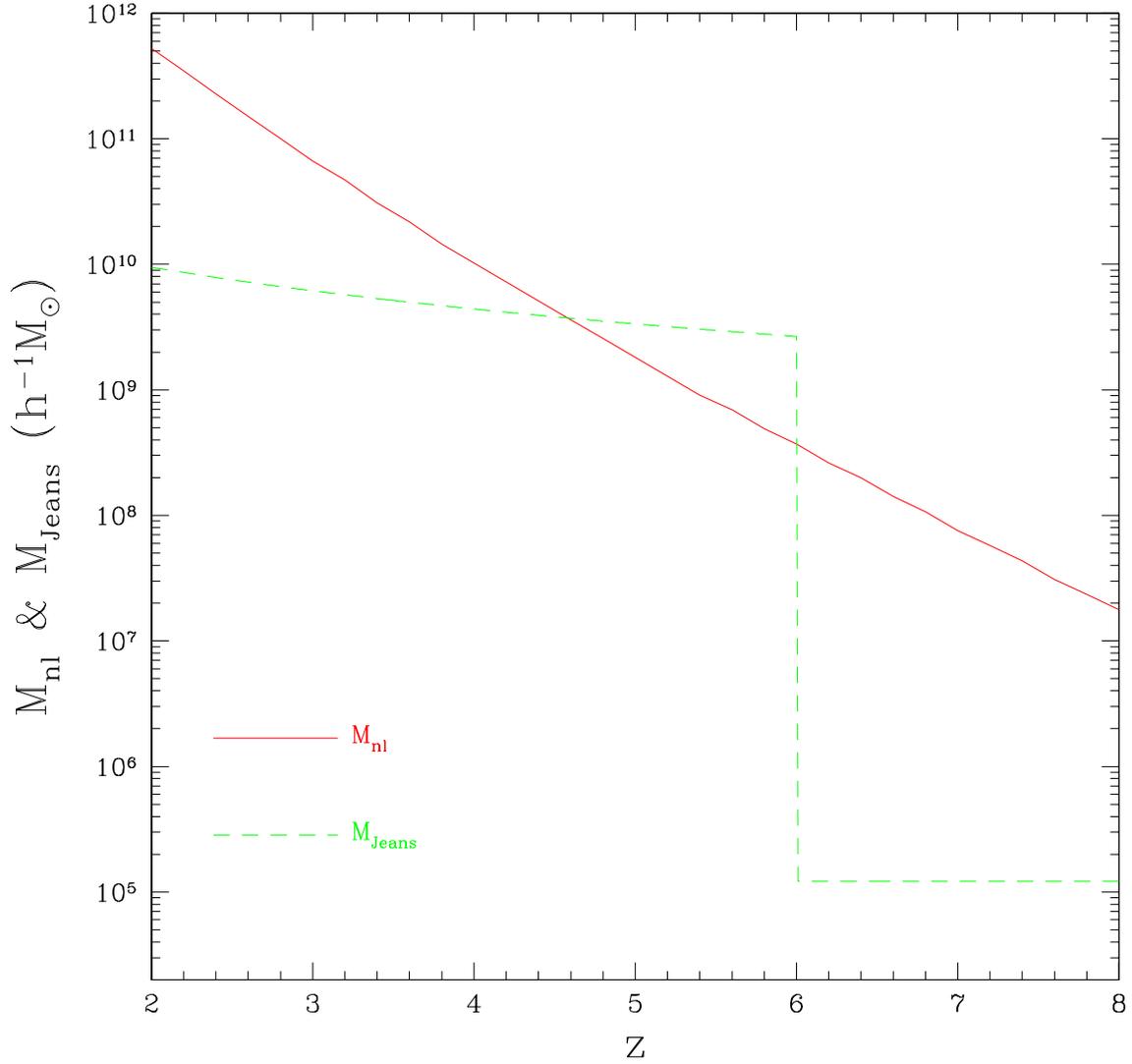}
 \caption{Nonlinear mass $M_{\rm nl}$ and linear Jeans mass $M_{\rm Jeans}$
 as a function of redshift, with the simplified assumption that
 the universe was instantaneously ionized at $z=6$.
 The intergalactic medium is assumed to have the same temperature
 as the cosmic microwave background at $z\ge 6$ and
 photo-heated to $1.5\times 10^4$Kelvin at $z<6$. A $\Lambda$CDM model
 with $\Omega=0.3$, $\Lambda=0.7$, $n=1$ and $\sigma_8=0.8$
 is assumed in order to compute $M_{\rm nl}$.
 }
 \label{nl}
 \end{figure}

 Figure \ref{nl} shows 
 the nonlinear mass $M_{\rm nl}$ and linear Jeans mass $M_{\rm Jeans}$
 as a function of redshift. 
 The nonlinear mass $M_{\rm nl}$ is the mass within a top-hat window
 at which the linear rms density fluctuation is unity,
 and 
 $M_{\rm Jeans}=1.5\times 10^{10} ({T\over 10^4{\rm K}})^{3/2} (1+z)^{-3/2}\Omega_M^{-1/2} h^{-1}\msun$ (Peebles 1993),
 where $T$ is the gas temperature
 and $\Omega_M$ is the present matter density parameter.
 The assumption of instantaneous reionization is 
 just for the sake of plotting convenience and is not required to make our point.
 In a real universe the reionization phase is complicated 
 and is thought to progress on a time scale of
 a Hubble time, during which the mean (volume weighted) radiation field builds up
 slowly up to a value of approximately
 $10^{-24}~$erg/cm$^2$/hz/sec/sr at Lyman limit,
 followed by a brief phase, when 
 the majority of the baryons are ionized and a sudden jump 
 in the amplitude of the mean radiation field intensity at Lyman limit
 to $10^{-22}-10^{-21}~$erg/cm$^2$/hz/sec/sr 
 occurs within a redshift interval of a fraction of unity 
 (Miralda-Escud\'e \etal 2000; Gnedin 2000a).
 We simply call this epoch of a sudden rise in radiation field
 the reionization epoch,
 marking the completion of the reionization process.

 We see, in Figure 2,
 that gas is able to accrete onto halos with mass greater than
 the Jeans mass $\sim 10^5\msun$ prior to $z\sim 6$.
 Stars formed inside 
 halos with $M=10^8-5\times 10^8\msun$
 are presumably primarily responsible for producing 
 most of the ionizing photons immediately prior to $z\sim 6$,
 where atomic line cooling provides an efficient cooling process.
 Right after reionization, there is 
 a dramatic inversion of the situation
 from $z\sim 6$ to $z\sim 4.6$, where Jeans mass surpasses the nonlinear mass.
 Such a switch has some interesting consequences:
 gas accretion onto the most abundant halos (about the nonlinear mass)
 is suddenly suppressed, simply because the ambient
 gas has a temperature that is higher than the virial temperature
 of the typical halos and is incapable of cooling to a lower temperature.
 Gas accretion and thus star formation may be brought to a halt in
 typical halos
 (larger halos with higher virial temperature 
 may still be able to form stars with a somewhat reduced rate 
 due to the same effect).
 Below $z\sim 4.6$, nonlinear mass again becomes larger than
 the Jeans mass and gas is once again capable of accreting
 onto typical halos with mass $\ge 4\times 10^9\msun$,
 and star formation rate bounces back.
 We note that the cosmological model adopted may not be
 correct in fine details and is used only to illustrate the process.
 Therefore, the exact crossing redshift of the two curves at $z=4.6$
 should not be taken at its face value, but rather as a generic feature.

 Let us now relate the star formation rate to the 
 ionizing background radiation field.
 Our treatment is intended to be simple but we think it contains 
 the essence of the relevant physical processes.
 For gas immersed in a meta-galactic ionizing radiation background,
 locally hydrogen photoionization is balanced by hydrogen recombination
 (to levels $n=2$ or higher):
 \begin{equation}
 \Gamma n_{HI} = n_{HII} n_e \alpha(T),
 \end{equation}
 \noindent
 where $\Gamma$ is the photoionization rate (see Figure 1),
 $n_{HI}$, $n_{HII}$ and $n_{e}$
 are the neutral hydrogen,
 ionized hydrogen,
 and electron number densities, respectively,
 and $\alpha(T)$ is the hydrogen recombination rate.
 Averaging spatially both sides of equation (1)
 gives 
 \begin{equation}
 <n_{HI}> = {C<n_{H}>^2 \alpha(T)\over \Gamma},
 \end{equation}
 \noindent
 where $C$ is the clumping factor of all gas participating 
 in receiving ionizing photons (i.e., excluding gas in optically
 shielded regions).
 For simplicity we have assumed that the
 gas is composed entirely of hydrogen.
 Since the neutral hydrogen fraction is much less than unity,
 we have simply replaced both the ionized hydrogen 
 and electron number densities by the total hydrogen number density $n_{H}$.
 The mean free path of an ionizing photon
 in comoving length units is then
 \begin{equation}
 \lambda = {1+z\over <n_{HI}>\sigma_H}={\Gamma (1+z)\over C(z)<n_H>^2\sigma_H\alpha},
 \end{equation}
 \noindent
 where $\sigma_H$ is the hydrogen photoionization cross-section.
 Using $\alpha=4\times 10^{-13} {\rm cm}^3 {\rm sec}^{-1}$,
 $\Gamma=1.5\times 10^{-13}~$sec$^{-1}$ (see Figure 1)
 and $\sigma_H=2\times 10^{-18}~$cm$^2$
 we obtain $\lambda$ in comoving megaparsecs:
 \begin{equation}
 \lambda = 74 C(z)^{-1}\left({\Omega_b h^2\over 0.02}\right)^{-2}\left({1+z\over 7}\right)^{-5} {\rm comoving\ Mpc}.
 \end{equation}
 \noindent
 Evidently, at $z \sim 6$,
 the mean free path of an ionizing photon
 is much smaller than the Hubble radius
 and radiative processes may be treated ``locally",
 by ignoring cosmological effects.
 But the mean free path is much greater than
 the typical separation between
 ionizing sources ($\sim 1~$comoving Mpc).
 In this case
 we may relate the mean {\it comoving} specific emissivity of ionizing radiation
 $\epsilon_\nu$ (in units of erg/sec/hz/comoving cm$^3$)
 to the mean ionizing radiation intensity $J_{\nu}$
 (in units of erg/sec/hz/sr/cm$^2$) by 
 \begin{eqnarray}
 J_{\nu} &= {\epsilon_\nu \lambda (1+z)^2\over 4\pi}  \nonumber\\
 &= 8.9\times 10^{26} C(z)^{-1}\epsilon_\nu ({\Omega_b h^2\over 0.02})^{-2} ({1+z\over 7})^{-3}.
 \end{eqnarray}
\noindent
%We examine closely the various terms in equation (5).
 The evolution of $C(z)$ is complicated and currently undetermined
 without detailed simulations that ionized the universe at $z\sim 6$.
 However, simple estimates may be made to gauge the range of variation.
 In the linear regime we have $C(z)\propto (1+z)^{-2}$
 and in the extreme nonlinear regime $C(z)\propto (1+z)^{0}$;
 the actual situation is likely to be between these two limiting
 cases, parameterized as $C(z)\propto (1+z)^{-2+\beta}$ with $\beta>0$.
 %After the reionization is complete, recombination is likely
 %dominated by dense regions in halos that are not shielded 
 %(Miralda-Escud\'e \etal 2001); in this case, $\beta$ may be close to zero.
 Inserting $C(z)$ into equation (5) yields
 \begin{equation}
 J_{\nu} \propto {\epsilon_\nu (1+z)^{-1-\beta}}
 \end{equation}
 \noindent
 with $\beta>0$.
 Clearly, in order for $J_\nu$ to increase with redshift,
 as indicated by observations in the redshift range
 $z=5.2-5.6$ as shown in Figure 1,
 the comoving specific emissivity or comoving star formation rate 
 has to increase with redshift.
 Taking the values shown in Figure 1 in the range $z=5.6-5.2$,
 we obtain $J_\nu\propto (1+z)^{9.8\pm 3.4}$. 
 (This fit used the 3rd
 through 5th filled squares in Figure \ref{Gev}.  We obtain a
 flatter slope, $5.3\pm2.1$, with an acceptable 
 $\chi^2$, if we add the 2nd point, at $z\simeq 5.1$.)
 This requires a decrease in star formation rate
 from $z=5.6$ to $z=5.2$ as
 $\epsilon \propto (1+z)^{10.8\pm 3.4 - \beta}$.
 Our analysis is in qualitative agreement with 
 Barkana \& Loeb (2000), who pointed out
 that the decrease in star formation subsequent to 
 cosmological reionization may be detectable 
 in the evolution of number counts of faint galaxies by NGST.
 It may be that we have detected this signature 
 in the evolution $J_\nu$ or $\Gamma$.

 \section{Conclusions}

 In hierarchical structure formation theory
 the evolution of the meta-galactic ionizing radiation background
 may be characterized by five distinct phases:
 1) from $z\sim {15-20}$ to $z_{ri}$ (which marks the end of reionization)
 the radiation field builds up slowly before individual ionizing HII 
 regions overlap (``pre-overlap period"; Gnedin 2000a)
 2) in a brief period up to $z_{ri}$ 
 the majority of baryons are ionized
 in a short time scale and the radiation field jumps up by
 about two to three orders of magnitude, completing the reionization process
 (``reionization epoch");
 3) from $z_{ri}$ to $z_{Jeans}$ the radiation field pulls back 
 significantly due to a decrease in star formation rate as a result
 of reionization (we call this ``aftermath of reionization");
 4) $z_{Jeans}$ to $z_{peak}\sim 1-2$ the radiation field rises
 steadily with contributions from both stars in larger
 galaxies and quasars;
 5) $z_{peak}\sim 1-2$ to $z=0$ the radiation field drops off sharply due to 
 the combined
 effect of decreasing star/quasar formation rate and cosmological
 effects.
 Consistent with theoretical calculations,
 observations have confirmed 
 phases (4) (e.g., McDonald \& Miralda-Escud\'e 2001;
 shown in Figure 1 of this paper) and (5) (e.g., Shull \etal 1999).

 In this paper we have interpreted the absorption in the
 spectra of the latest
 high redshift quasars in the redshift
 range $z=4.9-6.1$, searching for salient features pertaining to
 the ``reionization epoch" (phase 2)
 and the ``aftermath of reionization" (phase 3).
 Quite intriguingly,
 we appear to be seeing
 the ``aftermath of reionization" 
 and the end of the ``reionization epoch" 
 in these spectra,
 both of which seem to point to a cosmological reionization redshift
 of $z_{ri}\sim 6$.
 Specifically, a predicted decrease of star formation rate
 in the aftermath of the reionization is indicated
 by the observations.
 While a few more high redshift quasars are required to check that 
 the complete absorption at $z=6.1$ is not some kind of anomaly
 (e.g., a rare leftover neutral patch in an almost completely ionized
 universe),
 %While larger samples of high redshift quasars would be required
 %to check cosmic variance and substantially improve the signal-to-noise 
 %to make the determination of the ``reionization epoch" conclusive, 
 it appears that our conclusion with regard to
 the ``aftermath of reionization" is more secure due to
 %much increased signal-to-noise.
 its independence of any single spectrum.
 It is still very urgent to enlarge the observational data sets to 
 greatly firm up the conclusion and sharpen up the rate of evolution.
 If confirmed by future observations, this may be indication
 that our current understanding of galaxy formation may be
 approximately valid up to redshift $z\sim 6$.
 Furthermore, as suggested by Barkana \& Loeb (2000), 
 it will be important to 
 independently detect this ``aftermath of reionization" signature 
 in the evolution of number counts of faint galaxies by NGST.

 Since we appear to be observing the period directly after reionization,
 it becomes more important to perform dedicated simulations that
 achieve reionization at the appropriate redshift $z\sim 6$.
 This paper should be understood as an
 attempt to guide that effort by making some preliminary comparisons
 between data and an existing simulation in which reionization 
 occurred somewhat earlier.
 Such future simulations are quite demanding,
 because not only high resolutions are required but also
 an accurate treatment of three-dimensional radiative transfer
 must be implemented.

 \acknowledgments
 This research is supported in part by grants AST93-18185 and ASC97-40300.
 We thank Xiaohui Fan, Vijay Narayanan and David Weinberg 
 for useful discussion, and Nick Gnedin for his HPM code.

											    \end{document}